\documentclass[aps,pra,twocolumn,superscriptaddress,nofootinbib]{revtex4-2}
\usepackage{graphicx}
\usepackage{amssymb}
\usepackage{amsmath}
\usepackage{footnote}
\usepackage{bm}
\usepackage{xcolor}
\usepackage{verbatim}
\usepackage{kotex}
\usepackage{hyperref}
\hypersetup{
     colorlinks=true,
     linkcolor=blue,
     filecolor=blue,
     citecolor = black,
     urlcolor=black,
     }
\usepackage[all]{hypcap}

\begin{document}

\title{Half-quantum vortex generation in a two-component Bose-Einstein condensate by an oscillatory magnetic obstacle}

\author{Jong Heum Jung}
\affiliation{Department of Physics and Astronomy, Seoul National University, Seoul 08826, Korea}

\author{Y. Shin}
\email{yishin@snu.ac.kr}

\affiliation{Department of Physics and Astronomy, Seoul National University, Seoul 08826, Korea}

\affiliation{Institute of Applied Physics, Seoul National University, Seoul 08826, Korea}

\affiliation{Center for Correlated Electron Systems, Institute for Basic Science, Seoul 08826, Korea}

\date{\today}

\begin{abstract}
We numerically investigate the dynamics of vortex generation in a two-dimensional, two-component Bose-Einstein condensate subjected to an oscillatory magnetic obstacle. The obstacle creates both repulsive and attractive Gaussian potentials for the two symmetric spin-$\uparrow$ and $\downarrow$ components, respectively. We demonstrate that, as the oscillating frequency $f$ increases, two distinct critical dynamics arise in the generation of half-quantum vortices (HQVs) with different spin circulations. Spin-$\uparrow$ vortices are nucleated directly from the moving obstacle at low $f$, while spin-$\downarrow$ vortices are created at high $f$ by breaking a spin wave pulse in front of the obstacle. We find that vortex generation is suppressed for sufficiently weak obstacles, in agreement with recent experimental results by Kim {\it et al.} [Phys.~Rev.~Lett. {\bf 127}, 095302 (2021)]. This suppression is caused by the finite sweeping distance of the oscillating obstacle and the reduction in friction in a supersonic regime. Finally, we show that the characteristic length scale of the HQV generation dynamics is determined by the spin healing length of the system.

\end{abstract}

\maketitle

\section{Introduction}
\label{sec:intro}
 
A symmetric two-component Bose-Einstein condensate (BEC) was realized in recent experiments with $^{23}$Na in two hyperfine spin states~\cite{Raman11,Liu14,R20}. The BEC is a binary superfluid system with $\mathbb{Z}_2$ symmetry and represents a minimal setting to study spin superfluidity and related magnetic properties~\cite{Sonin10,Sonin19}. Many fascinating phenomena have been observed, including spin superflow behavior~\cite{R16,R19}, spin sound propagation~\cite{R26}, and novel topological objects such as half-quantum vortices (HQVs)~\cite{R20,R21} and magnetic solitons~\cite{R24,R25}. Recently, the critical dissipative dynamics of the two-component BEC was studied using an oscillating magnetic obstacle~\cite{JHK}. Notably, it was observed that HQVs were not generated for a weak magnetic obstacle; only spin waves were excited. Since HQVs involve both mass and spin circulations, it was speculated that HQVs require a magnetic obstacle of sufficient strength to generate sufficient mass flow perturbations necessary for their creation~\cite{JHK}.

Meanwhile, the oscillatory movements of an obstacle may affect critical conditions for vortex generation~\cite{Jackson, fujimoto2010, Fujimoto, reeves2012}. In a recent experiment with a single-component BEC, a similar observation was made that vortex dipole shedding was inhibited for an oscillating repulsive obstacle when the obstacle strength becomes too weak, and it was shown that the inhibition was due to the insufficient dragging time for the finite sweeping distance of the oscillating obstacle~\cite{YLim}. This result suggests that the suppression of HQV generation observed in \cite{JHK} might also be associated with the magnetic obstacle's oscillatory motion. The possibility is further conceivable because the characteristic length scale for HQVs in the two-component BEC is significantly larger than that for ordinary quantum vortices in the single-component BEC.

In this paper, we numerically investigate the HQV generation dynamics of an oscillatory magnetic obstacle in the two-component BEC and elucidate the vortex suppression mechanism for weak obstacles. We demonstrate two distinct generation dynamics for the two types of HQVs and attribute their differences to the obstacle's spin-dependent character. We determine the critical stirring frequencies as a function of the obstacle’s strength and find that HQV generation is suppressed below a threshold strength, consistent with the experimental observation in \cite{JHK}. From the comparison to the critical velocity of a uniformly moving obstacle and the oscillation amplitude dependence of the threshold obstacle strength, we show that the suppression of HQV generation results from the finite sweeping distance of the oscillating obstacle and the friction reduction in a supersonic regime. Additionally, we show that the spin healing length of the system dominantly governs the HQV generation dynamics. Our study provides a qualitative explanation of the previous experimental results~\cite{JHK} and highlights the influence of the finite sweeping distance of the obstacle on critical energy dissipation.

\section{Theoretical model}
\label{sec:model}

In the experiment of \cite{JHK}, a BEC of an equal mixture of two miscible components, denoted by spin-$\uparrow$ and $\downarrow$, was prepared with $^{23}$Na in the $|F=1, m_F=1\rangle$ and $|F=1,m_F=-1\rangle$ states. Their intracomponent interactions are identical, and the BEC comprises a symmetric binary superfluid system. A magnetic obstacle was formed by focusing a Gaussian laser beam, whose frequency was adjusted to provide repulsive and attractive potentials of the same magnitude to the spin-$\uparrow$ and $\downarrow$ components, respectively.

Because the BEC has highly oblate geometry, we assume that the BEC dynamics is effectively frozen for the tight confining, $z$ direction and investigate the vortex generation through numerical simulations of the two-dimensional (2D) coupled Gross–Piateviskii equations (GPEs),
\begin{equation}
    i\hbar \partial_t \Psi_i = \Big[ -\frac{\hbar^2}{2m}\nabla^{2}  + U_i(\textbf{r})  + g_0 \vert \Psi_i \vert ^2  +  g_{\uparrow \downarrow} \vert \Psi_j \vert ^2 -\mu \Big] \Psi_i,   
    \label{GPE}
\end{equation}
where $i,j \in \{\uparrow,\downarrow\}$, $i \neq j$, $\hbar$ is the reduced Planck constant, $m$ is the particle mass, and $\Psi_i$ is the macroscopic wave function of spin-$i$ component. $U_i(\textbf{r})$ is the external potential for spin-$i$ component, including the harmonic trapping potential $V_{\text{trap}}(\textbf{r})=\frac{1}{2}m(\omega_x^2 x^2 + \omega_y^2 y^2)$ and the magnetic obstacle potential $V_{\text{obs},i}({\bf r}) = s_{i} V_0\exp(-2\frac{\vert {\bf r}-{\bf r}_0 \vert^2}{\sigma^2})$ with $s_{\uparrow} = -s_{\downarrow} = 1$. Here, $V_0>0$ and ${\bf r}_0$ is the position of the obstacle. The third and fourth terms describe the intra- and inter-component interactions with coefficients $g_0$ and $g_{\uparrow\downarrow}$, respectively.
$\mu$ denotes the chemical potential of the BEC. In the ground state without the obstacle, the particle density at the trap center is $n_0=
\frac{2\mu}{g_0+g_{\uparrow\downarrow}}=2|\Psi_{\uparrow(\downarrow)}(0)|^2$. The total particle number is given by
$N=  \int \sum_i  \vert \Psi_i({\bf r},t) \vert^2 d{\bf r} = \frac{\pi}{2} n_0 R_x R_y$ with $R_{x(y)}=\sqrt{\frac{2\mu}{m\omega_{x(y)}^2}}$ being the Thomas-Fermi (TF) radius of the trapped BEC. 

\begin{figure}[t]
    \includegraphics[width=8.6cm]{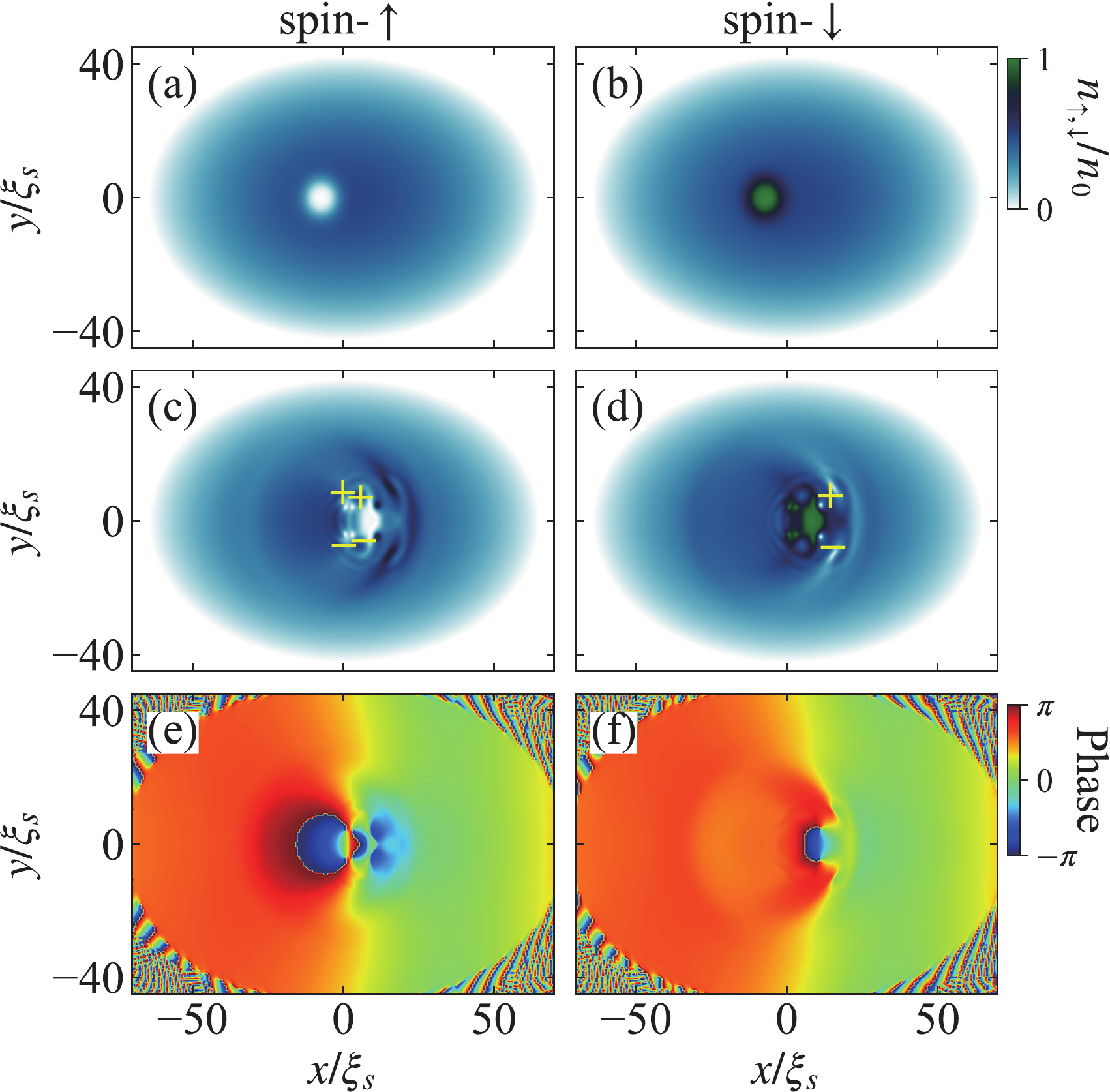}
    \centering
\caption{Two-component Bose-Einstein condensate (BEC) stirred by an oscillatory magnetic obstacle. The obstacle provides repulsive and attractive potentials to the spin-$\uparrow$ and $\downarrow$ components, respectively, and oscillates along $x$ direction in the center region of the harmonically trapped condensate. Density distributions $n_i(x,y)$ ($i\in\{\uparrow,\downarrow\}$) of the two spin components (a),(b) at the initial condition and (c),(d) after a half-period of oscillation. The obstacle strength is $V_0=1.5 V_s$ and the sweeping frequency is $f=f_s$. Vortices with counterclockwise(clockwise) mass circulation are denoted as $+$($-$). (e),(f) Phase distributions of the spin components, corresponding to (c),(d).}
    \label{explain}
\end{figure}

\begin{figure*}[t]
    \includegraphics[width=17.5cm]{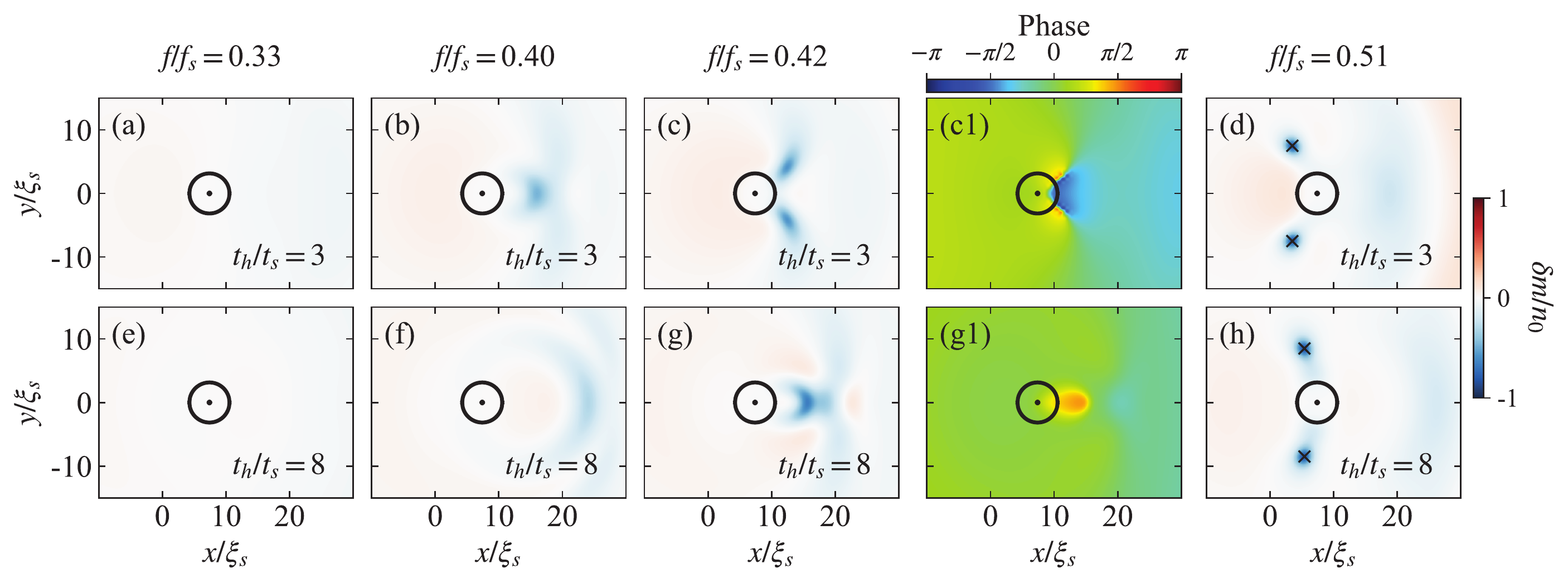}
    \centering
\caption{Spin-$\uparrow$ half-quantum vortex (HQV) generation by a magnetic obstacle in low $f$ regime. Magnetization distributions $\delta m (x,y)$ of the BEC for various sweeping frequencies $f$ (a)-(d) at $t_h=3t_s$ and (e)-(h) $t_h=8t_s$. $t_h$ is the hold time after finishing a half-period sweeping. The obstacle strength is $V_0/V_s=1.5$. (c1) and (g1) display the phase distributions of the spin-$\uparrow$ component in (c) and (g), respectively. The black circles with center dots indicate the impenetrable region ($V_{\text{obs},\uparrow}(\textbf{r}) > V_s$) for the obstacle. As $f$ increases, spin-$\uparrow$ HQVs are generated by the obstacle, which are indicated by $\times$ in (d) and (h).}
    \label{black excitation}
\end{figure*}

The two-component BEC system has two characteristic length scales: density and spin healing lengths, $\xi_n = \hbar/\sqrt{2m\mu}$ and $\xi_s = \gamma \xi_n$ with $\gamma = \sqrt{\frac{g_0+g_{\uparrow\downarrow}}{g_0-g_{\uparrow\downarrow}}}$, respectively. The corresponding time scales are given by $t_n = \hbar/\mu$ and $t_s = \gamma^2 t_n$, respectively. Using the change in variables, $\tau=t/t_s$ and ${\boldsymbol{\rho}}={\bf r} / \xi_s$, Eq.~(\ref{GPE}) is expressed in a dimensionless form as
\begin{eqnarray}
    i\partial_\tau \psi_i &=& \big[ -\nabla^{2}_{\boldsymbol{\rho}} + \tilde{U}_{i} + \nonumber \\
    && + (\gamma^2+1) \vert \psi_i \vert ^2  + (\gamma^2-1) \vert \psi_j \vert ^2 -\gamma^2  \big] \psi_i  \nonumber \\
    &=& \tilde{H}_i \psi_i
    \label{dimensionless_GPE}
\end{eqnarray}
with $\psi_i = n_0^{-1/2} \Psi_{i}$, $\nabla^2_{\boldsymbol{\rho}} = \xi_s^2 \nabla^2$, and $\tilde{U}_{i} = U_i/V_s$. Here, $V_s = \mu/\gamma^2 = m c_s^2$ is the spin energy scale, where $c_s$ is the speed of spin sound~\cite{R26}. When the peak potential $V_0$ of the obstacle is larger than $V_s$, the local chemical potential of the spin-$\uparrow$ component becomes negative at the position of the magnetic obstacle~\cite{JJH}.

We investigate the vortex generation dynamics by numerically calculating Eq.~(\ref{dimensionless_GPE}) for the experimental condition of \cite{JHK}; $\gamma =5.3$ for  $^{23}$Na~\cite{Knoop11}, $R_{x(y)}= 64.8 (42.4) \xi_s$, and the obstacle width $\sigma = 7\xi_s$. The position of the obstacle is controlled as ${\bf r}_0(t) = (-A\cos(2{\pi}ft),0)$ with $A = 7.4\xi_s$ and $f$ being the sweeping frequency. We define $f_s = \frac{c_s}{2\pi A}$, which denotes the sweeping frequency for which the obstacle's maximum speed is equal to $c_s$. The initial BEC state is set to be the ground state with a stationary obstacle at $x=-A$, which is obtained by the backward Euler pseudo spectral method using imaginary time propagation~\cite{GPELab1}. The system's evolution is calculated by a fourth-order time splitting pseudo spectral scheme~\cite{Bao,GPELab2}. The total system size is 160$\xi_s\times$120$\xi_s$ with 256$\times$256 grids. In Fig.~\ref{explain}, we display exemplary numerical results of the density and phase distributions of the two spin components before and after stirring the magnetic obstacle.

\section{Results and discussion}
\label{sec:result}

\subsection{Critical vortex shedding}
\label{subsec:A}

We consider an obstacle with $V_0 = 1.5 V_s$ and examine the BEC excitations for various sweeping frequencies after the first half period of the obstacle motion, i.e., a single stroke from $x=-A$ to $x=A$. The choice of the half-period time is to avoid the ensuing interactions between the obstacle and generated magnetic excitations. After the obstacle stops at $x=A$, we apply an additional hold time of $3t_s = 0.61 \sigma/c_s$, which is to let the vortices, if any, detached from the obstacle.

In Fig.~\hyperref[black excitation]{\ref{black excitation}}, a series of magnetization distributions, $\delta m=\delta n_{\uparrow}-\delta n_{\downarrow}$, is displayed for various $f$. Here, $\delta n_i$ denotes the deviation of the density distribution from the stationary state for the same external potential including that from the obstacle, and thus, $\delta m$ shows the magnetic excitations induced by the obstacle's motion. Spin wave excitations are generated by the obstacle even with very low $f=0.05f_s$. As $f$ increases, the system develops distinct magnetization distributions, and above a threshold frequency $f_{\uparrow, \textrm{HQV}}$, spin-$\uparrow$ HQVs are nucleated~[Fig.~\hyperref[black excitation]{\ref{black excitation}(d)}], which are identified with the phase singularities in the spin-$\uparrow$ component. Below $f_{\uparrow,\textrm{HQV}}$, as a precursor of vortex generation, a rarefaction-like pulse (RP) is created, which is a superposition of spin-$\uparrow$ rarefaction and spin-$\downarrow$ compression pulses [Fig.~\hyperref[black excitation]{\ref{black excitation}(b)}]. When $f$ approaches the critical frequency, the RP bifurcates along the $y$ direction, orthogonal to the propagation direction [Fig.~\hyperref[black excitation]{\ref{black excitation}(c)}] and later evolves into a pair of 
HQVs [Fig.~\hyperref[black excitation]{\ref{black excitation}(d)}]. 

\begin{figure*}[t]
    \includegraphics[width=17.5cm]{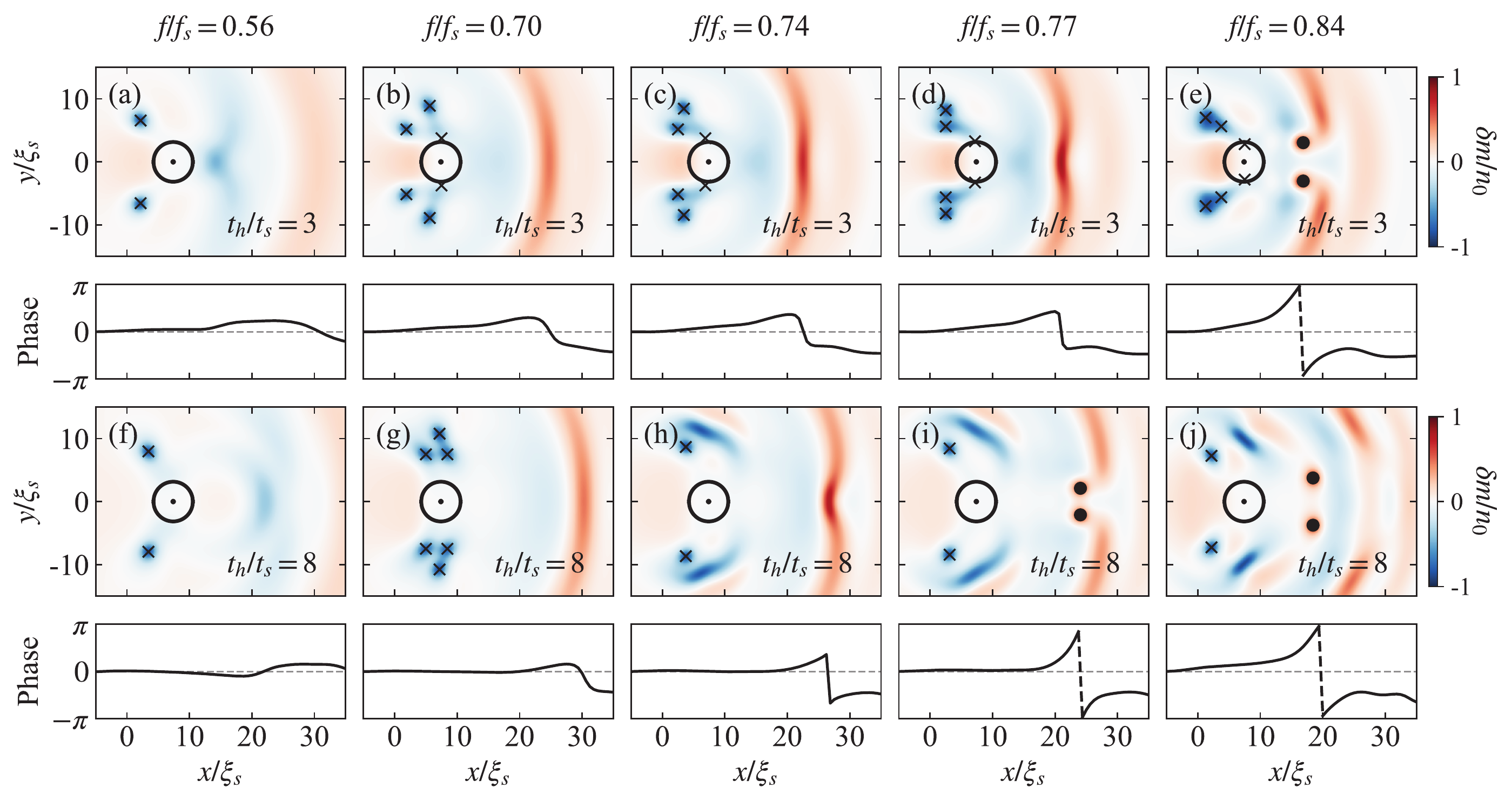}
    \centering
\caption{Spin-$\downarrow$ HQV generation by a magnetic obstacle in high $f$ regime. Magnetization distributions $\delta m (x,y)$ of the BEC for various $f$ (a)-(e) at $t_h=3t_s$ and (f)-(j) $t_h=8t_s$. $V_0/V_s=1.5$. The black circles with center dots indicate the boundary of the impenetrable region for the obstacle. Phase profiles of the spin-$\downarrow$ component along the $y=0$ line are displayed in the bottom of the magnetization images. A spin wave pulse develops in front of the obstacle and ruptures to generate a pair of spin-$\uparrow$ HQVs. Spin-$\uparrow$($\downarrow$) HQVs are marked as $\times$ ($\bullet$).}
    \label{white property}
\end{figure*}

In the bottom row of Fig.~\ref{black excitation}, numerical data for a longer hold time of $8t_s$ are displayed. In general, spin wave excitations propagate away from the obstacle. The RP generated at $f<f_{\uparrow,\textrm{HQV}}$ traverses the condensate with speed $c_s$, spreading out its magnetization [Fig.~\hyperref[black excitation]{\ref{black excitation}(f)}]. Meanwhile, the bifurcated RPs shown in Fig.~\hyperref[black excitation]{\ref{black excitation}(c)} merge into a single RP during propagation [Fig.~\hyperref[black excitation]{\ref{black excitation}(g)}], which might be attributed to the attraction by the magnetic obstacle. For $f>f_{\uparrow,\textrm{HQV}}$, a spin-$\uparrow$ HQV pair is stably formed and discharged from the obstacle, maintaining their magnetized cores [Fig.~\hyperref[black excitation]{\ref{black excitation}(h)}].

In Fig.~\ref{white property}, magnetization distributions for a further increase in $f$  are displayed. Notably, at high $f \geq 0.7 f_s$, we observe that a solitary spin wave (SSW) with $\delta m>0$ is generated in front of the obstacle, which preserves its magnetization structure during propagation [Figs.~\hyperref[white property]{\ref{white property}(b) and (g)}]. The magnetization at the core of the SSW is further enhanced with increasing $f$ and above a threshold frequency $f_{\downarrow,\text{HQV}}$, the SSW ruptures to generate a pair of vortices that have circulations of the spin-$\downarrow$ component [Fig.~\hyperref[white property]{\ref{white property}(e)}]. The SSW splitting time slightly depends on the sweeping frequency and in this study, we determine the characteristic value of $f_{\downarrow,\text{HQV}}$ from the BEC state at a hold time of $3 t_s$ after finishing the obstacle sweeping.

Our numerical results demonstrate two distinct vortex generation dynamics for spin-$\uparrow$ vortices at low $f$ and for spin-$\downarrow$ vortices at high $f$, respectively. The two processes are qualitatively different in the way of nucleating phase singularities in the corresponding spin component. The difference originates from the magnetic obstacle's spin-dependent character and in a simplified viewpoint, the two processes can be understood as vortex generation with repulsive and attractive obstacles, respectively. In the case of a repulsive obstacle, the particle density is depleted by the obstacle. Vortices are nucleated in the density-depleted region and shed directly from the moving obstacle; this corresponds to the spin-$\uparrow$ component which feels the magnetic obstacle repulsive. On the other hand, in the attractive obstacle case, rarefaction develops in front of the moving obstacle and once it is intensified above a threshold level, the rarefaction pulse splits into a vortex pair~\cite{Jackson,Aioi}; this corresponds to the spin-$\downarrow$ component that feels the magnetic obstacle attractive. An attractive obstacle typically exhibits a higher critical velocity than a repulsive one, which explains the observation of $f_{\uparrow,\textrm{HQV}} < f_{\downarrow,\textrm{HQV}}$.

\subsection{Excitation phase diagram}
\label{subsec:B}

\begin{figure}[t!]
    \includegraphics[width=8.6cm]{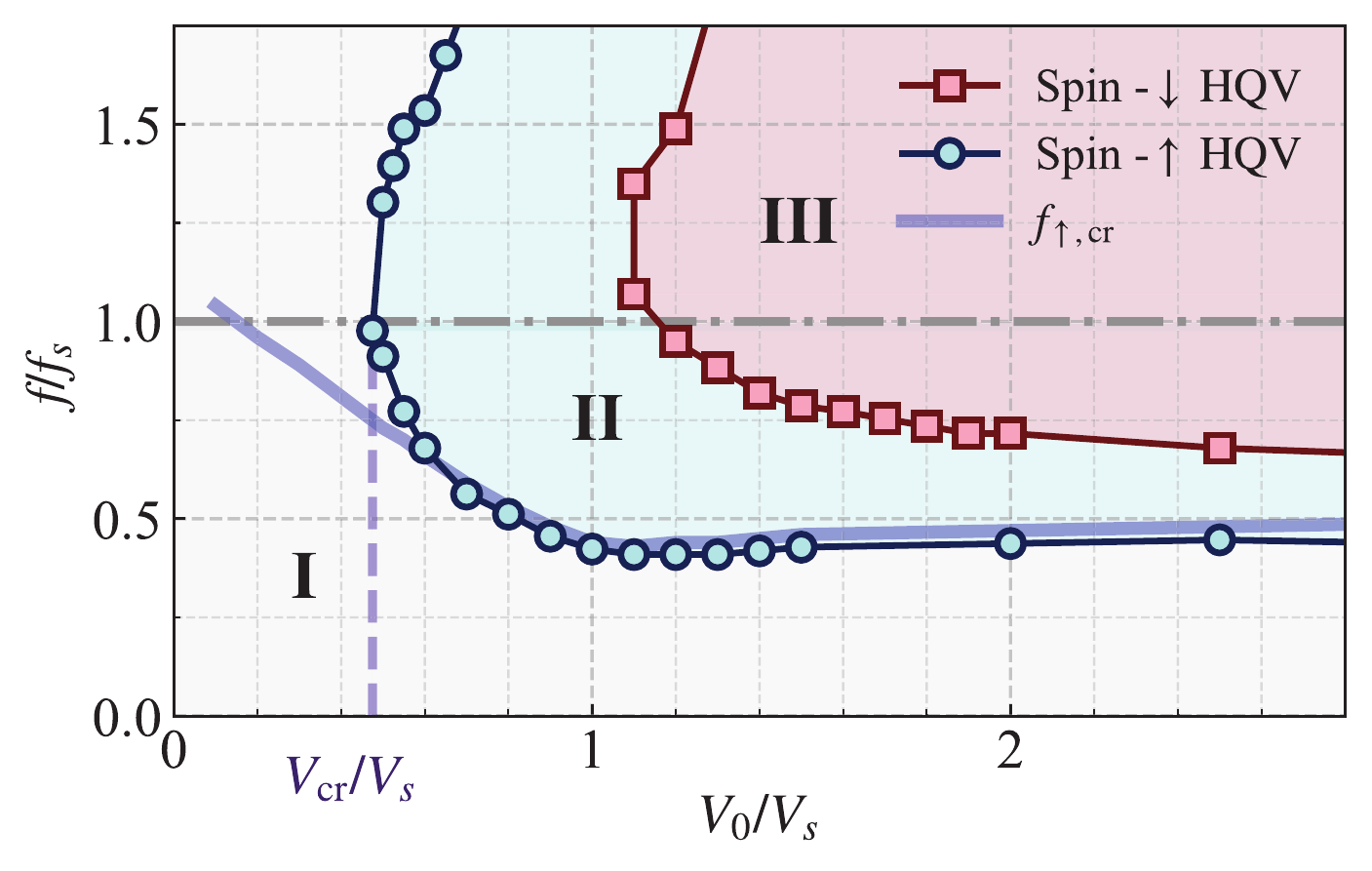}
    \centering
\caption{Excitation phase diagram in the $V_0$--$f$ plane. Three phases are identified: phase {\bf I} with no HQVs (white), phase {\bf II} with only spin-$\uparrow$ HQVs (blue-shaded), and phase {\bf III} with both types of HQVs (red-shaded). Markers are the critical frequencies numerically obtained for the corresponding phases. The gray dashed-dotted line indicates $f=f_s$. The purple dashed line represents the critical obstacle strength $V_\textrm{cr}$ for spin-$\uparrow$ HQV. The bright blue, thick line indicates $f_{\uparrow,\textrm{cr}} = \frac{1}{2\pi A} v_{\uparrow,\textrm{cr}}$, where $v_{\uparrow,\textrm{cr}}$ is the critical velocity of the magnetic obstacle that moves uniformly in a homogeneous BEC.}
    \label{phasediagram}
\end{figure}

In Fig.~\ref{phasediagram}, we plot the excitation phase diagram of the two-component BEC in the parameter space spanned by obstacle strength $V_0$ and oscillation frequency $f$ for the half-period motion of the obstacle. Three phases are indicated: phase {\bf I} without vortices, phase {\bf II} with only spin-$\uparrow$ HQVs, and phase {\bf III} with both types of HQVs. The critical frequencies, $f_{\uparrow,\textrm{HQV}}$ and $f_{\downarrow,\textrm{HQV}}$, determine the lower bounds in $f$ for phase {\bf II} and phase {\bf III}, respectively. Notably, each phase exhibits an upper critical frequency for a given $V_0$. Furthermore, there is a lower bound in $V_0$ for the phase. This means that when the obstacle strength is lower than the bound, the obstacle cannot produce the corresponding HQV, which is consistent with the experimental observation in Ref.~\cite{JHK}. We denote the critical obstacle strength for phase {\bf II} as $V_{\text{cr}}$, which represents the minimum $V_0$ to generate HQVs in the system. In our numerical study for the half-period obstacle motion, $V_{\text{cr}}=0.48V_s$.\footnote[1]{We also investigated a situation where small Gaussian white noises were added in the initial state to break the left-right symmetry with respect to the obstacle's moving direction. The excitation phase diagram was almost identical to that presented in Fig.~\ref{phasediagram}.}

For comparison, we calculate the critical velocity, $v_{\uparrow,\textrm{cr}}$, of the same magnetic obstacle for its spin-$\uparrow$ HQV nucleation with a constant linear motion in a homogeneous BEC. To determine $v_{\uparrow,\textrm{cr}}$, we examine the existence of a stationary solution for the system for a given obstacle, using the imaginary time propagation method~\cite{Huepe}. The numerical result is plotted together in the phase diagram in terms of the representative frequency $f_{\uparrow,\textrm{cr}}=\frac{1}{2\pi A}v_{\uparrow,\textrm{cr}}$, i.e., the oscillating frequency to have the maximum velocity to be $v_{\uparrow,\textrm{cr}}$. In contrast with $f_{\uparrow,\textrm{HQV}}$, $v_{\uparrow,\textrm{cr}}$ is well determined for $V_0<V_\textrm{cr}$.

This observation suggests that the emergence of the critical strength in the half-period dragged obstacle is the consequence of two effects: finite distance sweeping and friction reduction in the supersonic regime. In vortex creation, a moving obstacle accumulates energy by changing the flow pattern around it, which takes a certain amount of time. Insufficient energy or sweeping time would result in the emission of intermediate products such as RP or SSW, as observed for $f<f_{\uparrow(\downarrow),\textrm{HQV}}$. Therefore, to finish the vortex formation within a given sweeping distance, the drag force by the obstacle needs to be stronger, which might be fulfilled by increasing the obstacle speed~\cite{Jackson}. However, when the obstacle enters the supersonic regime, i.e., moves faster than the speed of sound, the enhancement of drag force cannot be achieved by an increase in velocity. When the obstacle oscillates too fast, particles would feel the time-averaged potential of the obstacle, thus, reducing effective friction. This effect is indeed the origin of the upper critical frequency for vortex generation. In previous studies, it was shown that vortex excitation is inhibited for supersonic obstacle~\cite{Radouani,Engels,Pinsker}. We observe that the critical frequency at the critical strength $V_{\textrm{cr}}$ is close to $f_s$, corresponding to the speed of spin sound, which supports our description of $V_{\textrm{cr}}$.

\begin{figure}[t]
    \includegraphics[width=8.6cm]{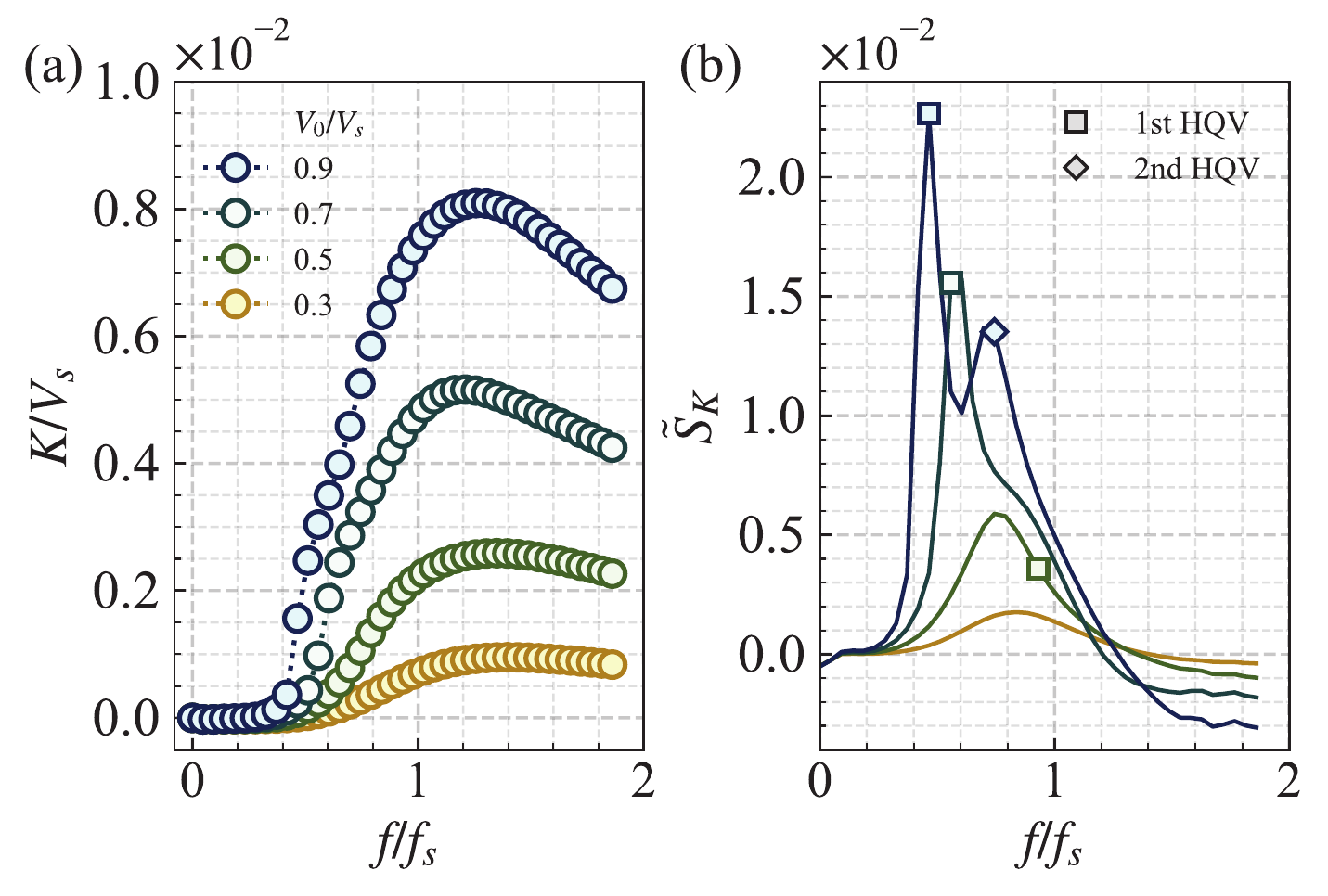}
    \centering
\caption{Onset of spin excitations. (a) Kinetic energy per particle, $K$ of the stirred BEC as a function of $f$ for various $V_0$ and (b) the normalized increasing rate $\tilde{S}_K$ = $\frac{f_s}{V_s}\frac{d K}{d f}$. The line color in (b) indicates the value of $V_0$ in (a). Squares and diamonds indicates the onset frequencies for the first and second spin-$\uparrow$ HQV pair generations, respectively.}           
\label{kinetic}
\end{figure}

In the experiment of \cite{JHK}, for a weak obstacle, although no HQVs were generated, the onset behavior of spin excitations was observed as the oscillating frequency increased. In our numerical simulations, we also observe such behavior for $V_0<V_{\text{cr}}$ and find that it is associated with the RP generation. In Fig.~\hyperref[kinetic]{\ref{kinetic}}, we display the kinetic energy per particle, $K$ after the half-period sweeping and its normalized increasing rate $\tilde{S}_K=\frac{f_s}{V_s}\frac{dK}{df}$ as functions of $f$ for various $V_0$. Here $K$ is calculated as 
\begin{equation}
K = V_s\frac{n_0\xi_s^2}{N} \int \Big( \vert \nabla_{\boldsymbol{\rho}} \psi_{\uparrow} \vert^2 +\vert \nabla_{\boldsymbol{\rho}} \psi_{\downarrow} \vert^2 \Big) d^2 \boldsymbol{\rho}
%{\int \Big( \vert \psi_{\uparrow} \vert^2 +\vert \psi_{\downarrow} \vert^2 \Big) d^2 \boldsymbol{\rho}}
    \label{k}.    
\end{equation}
The kinetic energy exhibits threshold behavior with increasing $f$ even for $V_0<V_\text{cr}$ and the initial increase of $\tilde{S}_K(f)$ is correlated with RP emission. For $V_0>V_\text{cr}$, the peak positions in $\tilde{S}_K(f)$ correspond to the critical frequencies $f_{\uparrow(\downarrow),\textrm{HQV}}$ for HQV generation. In the deep supersonic regime for $f>1.4f_s$, $K$ decreases with $\tilde{S}_K<0$, indicating friction reduction.

\begin{figure}[t]
    \includegraphics[width=8.6cm]{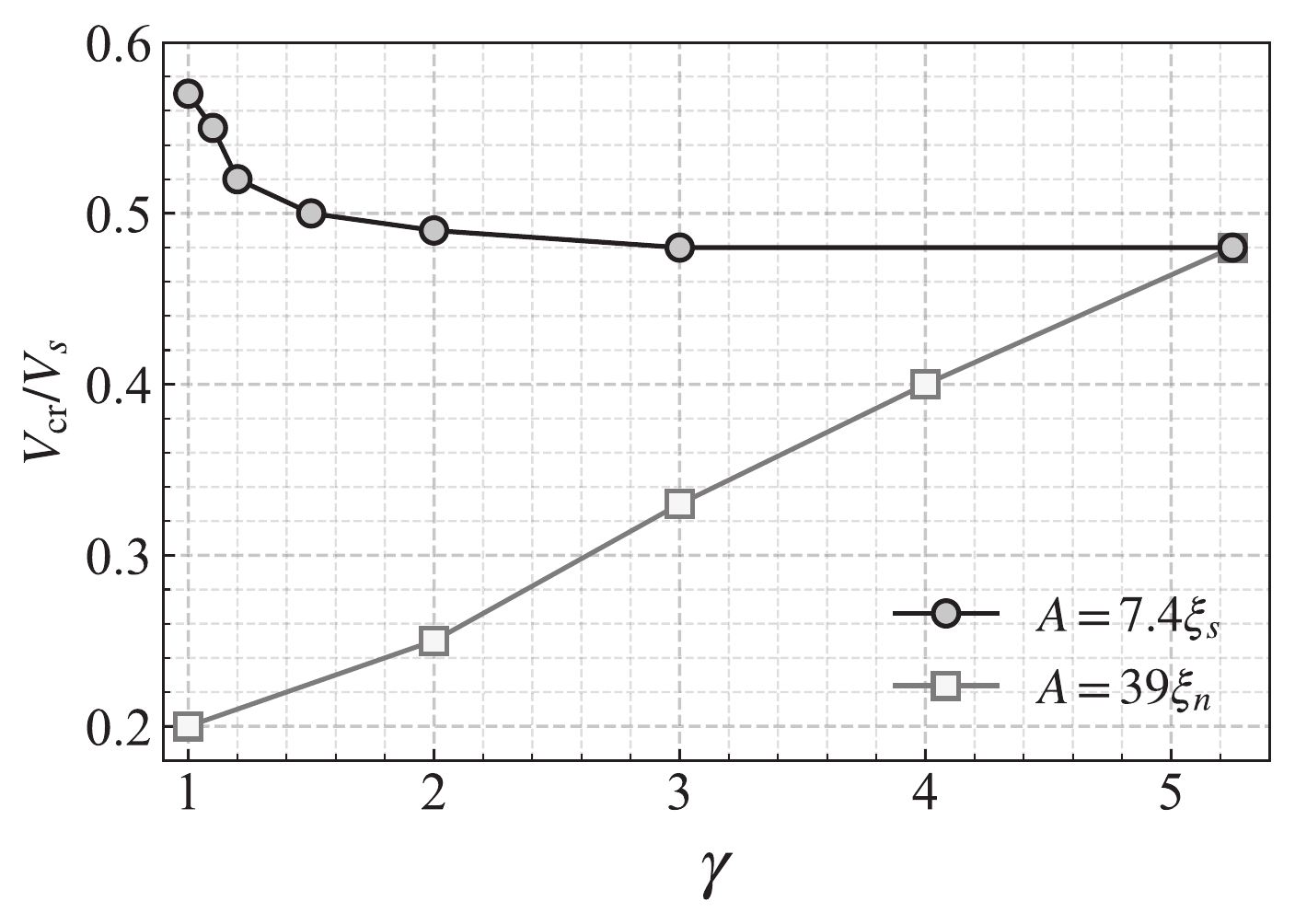}
    \centering
\caption{Dependence of the critical obstacle strength $V_{\textrm{cr}}$ on the interaction parameter $\gamma=\xi_s/\xi_n$. $V_{\text{cr}}$ are obtained for two sweeping distances: $A=7.4\xi_s$ (circles) and $39\xi_n$ (squares), which are equal for $\gamma=5.3$.}
    \label{compare}
\end{figure}

\subsection{Intercomponent interaction effect}
\label{subsec:D}

To understand the role of the intercomponent interactions in the HQV generation dynamics, we investigate the dependence of the critical strength $V_\textrm{cr}$ on the interaction parameter $\gamma=\xi_s/\xi_n$. We consider two different cases with varying $\gamma$, where the sweeping distance and the sample radii are kept fixed in units of either of $\xi_s$ (case \textbf{A}) or $\xi_n$ (case \textbf{B}), i.e., in \textbf{A}, $A=7.4\xi_s$ and $(R_x, R_y)=(64.8, 42.4)\xi_s$ and in \textbf{B}, $A=39\xi_n$ and $(R_x, R_y)=(343, 225)\xi_n$. For $\gamma=5.3$ as in the experiment, the two cases are identical to each other. The obstacle size is maintained to be $\sigma=7\xi_s$. 

\begin{figure*}[t]
    \includegraphics[width=16cm]{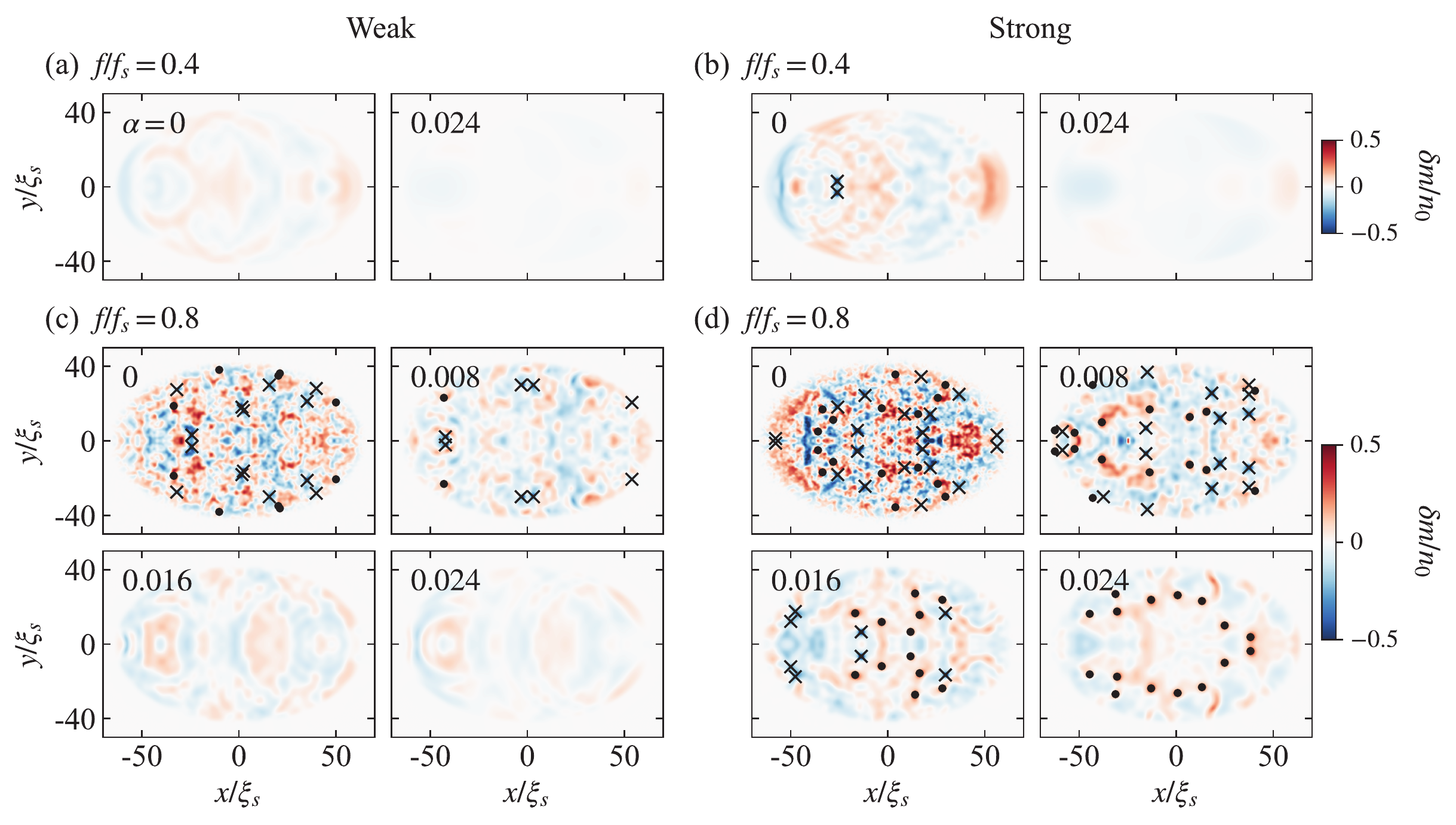}
    \centering
\caption{Damping effects in the HQV generation. Magnetization distributions $\delta m(x,y)$ of the BEC after multiple oscillations of the magnetic obstacle are calculated using Eq.~(\ref{dGPE}) for various damping parameters $\alpha$. The stirring time is $5/f_s$ (see the text for the stirring sequence). (a),(b) $f/f_s=0.4$ and (c),(d) 0.8. (a),(c) $V_0/V_s=0.9$ and (b),(d) 2.2. Spin-$\uparrow$($\downarrow$) HQVs are indicated by $\times$($\bullet$).}
    \label{nhqv}
\end{figure*}

In Fig.~\hyperref[compare]{\ref{compare}}, we display the evolution of $V_{\text{cr}}$ with $\gamma$ ranging from 1 to 5.3 for case \textbf{A} and \textbf{B}. Here $\gamma=1$ represents a system with no intercomponent interactions. In case \textbf{A}, $V_{\text{cr}}/V_s$ is almost insensitive to $\gamma$ $(>1.5)$, whereas in case \textbf{B}, $V_{\text{cr}}/V_s$ linearly increases from 0.20 to 0.48 as $\gamma$ increases from 1 to 5.3. These results indicate that the spin healing length dominantly governs the HQV generation dynamics in the two-component BEC. It is understandable because the core size of HQVs is also determined by $\xi_s$~\cite{Ji2008,Eto2011}. Note that in case \textbf{B}, the sweeping distance of the obstacle effectively increases with decreasing $\gamma$ as $A/\xi_s=39/\gamma$, thus, resulting in a decrease of $V_{\text{cr}}$. This confirms the finite sweeping distance effect that was discussed as the origin of the appearance of $V_\text{cr}$.

To understand the insensitivity of $V_\textrm{cr}$ to $\gamma$ in case \textbf{A}, we rewrite the time evolution operator in Eq.~(\ref{dimensionless_GPE}) as
\begin{eqnarray}
    \tilde{H}_i &=& -\nabla^{2}_{\boldsymbol{\rho}} + \tilde{U}_{i} + \vert \psi_i \vert ^2  - \vert \psi_j \vert ^2 + \gamma^2 \delta \tilde{n},
    \label{dimensionless_GPE2}
\end{eqnarray}
where $\delta \tilde{n} = \vert \psi_i \vert ^2+ \vert \psi_j \vert ^2 -1$ is the rescaled density variation of the BEC. Because the spin energy scale $(V_s)$ of the system is $\gamma^2$ times smaller than the density energy scale $(\mu)$, it is reasonable to expect that spin perturbations would be accompanied by $\gamma^{2}$ times smaller density perturbations, i.e., $\gamma^2 \delta \tilde{n} = \mathcal{O}(\gamma^{0})$. Then, Eq.~(\ref{dimensionless_GPE2}) is approximately independent of $\gamma$ for $\gamma \gg 1$, explaining the uniformity of $V_\textrm{cr}$ for $\gamma > 1.5$ in case \textbf{A}. Similar discussions on the separation of spin and mass sectors were provided for describing the spatial structures of magnetic solitons~\cite{Qu} and HQVs~\cite{Kasamatsu}.

\subsection{Damping effect}
\label{subsec:C}

In the experiment of \cite{JHK}, the obstacle oscillated for 1~s, which is about five oscillation periods for $f=f_s$. Now we perform numerical simulations of the experiment following its full sequence: preparing a stationary BEC, turning on the obstacle for 300~ms ($\approx 53t_s$), oscillating it for 1~s, and then turning off the obstacle for 300~ms. The numerical results of the magnetization distribution are displayed in Fig.~\ref{nhqv} ($\alpha=0$), where $V_0/V_s=0.9$ and $2.2$, which were the obstacle strengths employed in the experiment, and $f/f_s=0.4$ and $0.8$.
HQVs are more easily generated with multiple oscillations of the obstacle in comparison with the half-period sweeping case. Spin-$\uparrow$ HQVs are produced by the strong obstacle even when $f$ is smaller than $f_{\uparrow,\text{HQV}}(=0.44f_s)$ [Fig.~\hyperref[nhqv]{\ref{nhqv}(b)}]. Furthermore, spin-$\downarrow$ HQVs are also generated by the weak obstacle [Fig.~\hyperref[nhqv]{\ref{nhqv}(c)}], which is different from the excitation phase diagram in Fig.~\ref{phasediagram}. From the inspection of the time evolution of the magnetization distribution during the obstacle's oscillations, we find that spin excitations induced by the moving obstacle reflect from the condensate boundary and propagates back to the trap center region, thus, facilitating the generation of HQVs. The interactions between the obstacle and generated spin excitations are important in the multiple oscillation case.

In a realistic experimental condition, the sample temperature is not low enough, particularly, in the spin energy scale. Thus, the sample is expected to possess a sensible magnitude of thermal spin fluctuations, which would attenuate magnetic waves and suppress vortex generation. As the simplest approach to include the finite temperature effect, we employ the damped GPE of the binary system, which is given by~\cite{Achilleos}
\begin{equation}
    i\partial_\tau \psi_{i} = (1-i\alpha) \tilde{H}_i \psi_{i}
     \label{dGPE}
\end{equation}
with a damping parameter $\alpha$. In Fig.~\ref{nhqv}, we display the numerical results with the same multiple oscillation sequences for various $\alpha$. It is clear that spin wave and vortex generations are suppressed as $\alpha$ increases. We find that $\alpha \geq 0.024$ is required to fully suppress the HQV generation for the weak obstacle with $V_0/V_s=0.9$ [Fig.~\ref{dGPE}(c)] as observed in \cite{JHK}. We note that $\alpha\sim10^{-2}$ is a characteristic value suggested in previous numerical studies~\cite{Achilleos}.  

An interesting observation with the damping effect is that spin-$\uparrow$ HQV pairs are annihilated in the interaction with the obstacle. As the obstacle changes its direction during oscillations, a spin-$\uparrow$ HQV pair generated by the obstacle may collide with the returning obstacle. The repulsive obstacle attracts the counter-propagating vortex pair~\cite{fujimoto2010, Fujimoto} and the vortex pair can be annihilated during the collision due to the damping effect. Conversely, spin-$\downarrow$ HQV pairs do not collide with the obstacle because they are created remotely from the obstacle. Because of this difference, in the strong obstacle case with $\alpha=0.024$, we observe that spin-$\downarrow$ HQVs dominantly populate after stirring by the obstacle [Fig.~\hyperref[nhqv]{\ref{nhqv}(d)}]. However, we need to point out that in the experiment, the numbers of HQVs of both types were well balanced for the strong obstacle, hinting the limitations of our numerical approach including thermal fluctuations and damping effects.

\section{Conclusion}
\label{sec:summary}

We have numerically investigated the dynamics of HQV generation by an oscillatory magnetic obstacle in a symmetric two-component BEC  using 2D GPE. Our results revealed two distinct excitation dynamics for spin-$\uparrow$ and $\downarrow$ HQVs in low and high $f$ regimes, respectively, which are characterized by vortex generation for effective repulsive and attractive obstacles. We observed the suppression of HQV generation at reduced obstacle strength, consistent with the experimental findings of Kim {\it et al.}~\cite{JHK}, and attributed it to the finite sweeping distance of the oscillatory obstacle and the reduction of the dragging force in the supersonic regime. Furthermore, we demonstrated that the HQV generation dynamics are dominantly governed by the length and time scales given by the spin interaction energy. Moreover, we highlighted the significant role of damping in the experiment. 

To extend our investigation of the dissipative dynamics of the binary superfluid system, we suggest exploring magnetic obstacles with different spatial structures and modifying the surrounding flow conditions. For instance, applying multiple laser beams with different frequencies could allow for the engineering of novel magnetic obstacles and new avenues for experimental exploration. In a recent numerical study, a magnetic obstacle potential was proposed to generate skyrmions~\cite{skyrmion}. Additionally, a magnetic obstacle could be placed under spin currents, instead of mass currents. By applying a magnetic field gradient, spin currents can be readily induced in an experiment~\cite{R16,R19}.

\begin{acknowledgments}
We thank Haneul Kwak for assistance in numerical calculations. This work was supported by the National Research Foundation of Korea (NRF-2018R1A2B3003373, NRF-2019M3E4A1080400) and the Institute for Basic Science in Korea (IBS-R009-D1).
\end{acknowledgments}

\end{document}